# Contact Binary Stars as Standard Candles

Slavek Rucinski[1,2]

*David Dunlap Observatory, University of Toronto, P.O. Box 360, Richmond Hill, Ontario, Canada L4C 4Y6*

**Abstract.** The orbital period is a measure of the size of a contact binary star, and can be combined with color to predict its luminosity to a fraction of stellar magnitude. This novel application of contact binary systems currently has limitations that we describe in the text but is potentially capable of providing a reasonable means of estimating distances to stellar systems at the level comparable to that achievable for Cepheids or other pulsating stars.

## 1. Introduction

For a sequence of topologically similar contact binaries along the Main Sequence, the orbital period is a simple measure of their sizes. In addition, it is a distance-independent parameter, so it is insensitive to reddening and its local peculiarities. Realization of this advantage of the orbital period over all other parameters of contact binary stars is really the background of ideas presented here.

Contact binaries show a wide range of effective temperatures and – for a given temperature – a considerable range of periods can be observed due to evolution within the Main Sequence band. Thus, to the lowest order, luminosities of contact binaries can be described as dependent on two quantities: the orbital period, $P$ (hereinafter expressed in days), and a measure of $T_{eff}$, usually an effective-temperature sensitive color. Reduction to only two parameters is a strong simplification, but it must work, especially for faint systems in distant stellar clusters and for by-products of gravitational micro-lensing searches where no other information is available. We will return to these details further. Now we only note that we are witnessing a tremendous expansion in numbers of contact binaries from about 600 listed in the current variable-star catalogues to several thousands (perhaps of the order of 10,000 or more) recently discovered during the micro-lensing projects. These new systems urgently need characterization and beg to be utilized for various astrophysical purposes.

This paper is a summary of recent contributions aimed at establishing simple and efficient techniques of extracting useful information for large numbers of contact binaries. While the luminosity calibration for ordinary, solar-abundance systems and for metal-poor systems have been separately discussed (Rucinski

---

[1]rucinski@astro.utoronto.ca

[2]Affiliated also with the Department of Physics and Astronomy, York University, Toronto

1994b = CAL1 and Rucinski 1995a = CAL2), a useful background might be
found in papers discussing constraints on the period–color relation for normal
(Rucinski 1992) and for metal-poor systems (Rucinski 1994a). The first attempt to apply the calibrations to by-product discoveries of contact binaries in
the OGLE micro-lensing project (Udalski et al. 1994) is described in Rucinski
(1995b = BWC). A simple technique of light curve characterization has been
presented in Rucinski (1993a). The work summarized here is continuing so this
review is just a progress report, and it is very far from a final word on the matter.

## 2. The basis of the calibration

As has been shown in CAL1, the absolute magnitude of a system consisting of
components with masses $M_1$ and $M_2$ ($q = M_2/M_1 \leq 1$), observed at maximum
light should depend on the system parameters as:

$$M_V = -10 \log T_{eff} + B.C.(T_{eff}) - \frac{10}{3} \log P$$

$$-\frac{5}{3} \log M_1 - \frac{5}{3} \log(1+q) - 2.5 \log S(q,f) + const$$

(1)

where $S(q,f)$ is the area of the common contact surface with the distance between mass centers used as a unit of length. $S$ is very weakly dependent on
$q$ but does depended on the degree-of-contact, $0 < f < 1$. There are however
indications that the contact is similar for most systems and probably quite weak
(Rucinski 1985).

We note that Eq. (1) is a purely geometrical relation, linking brightness
with emitting area and surface brightness. *It does not involve mass-luminosity
relation,* i.e. it is absolutely independent of the way a system became a contact
one or whether it might hide evolved stellar cores.

Our simplification consists of dropping all terms besides one dependent on
$T_{eff}$ and one dependent on $P$. The basis for this drastic step are indications
(still poorly documented) that more massive systems tend to have smaller $q$;
in addition, on the Main Sequence, more massive systems are hotter so that
the $M_1 = M_1(T_{eff})$ correlation should be quite obvious. The degree of contact for most systems seems to be weak, or might again correlate with the
effective temperature (Rucinski 1985, 1993b). In any case, we disregard all
these partial correlations and consider only simplified relations of the form
$M_V = M_V(log P, log T_e)$ which, we hope, will absorb all other relationships.
In practice, we consider: $M_V = a_P \times log P + a_C \times \text{color} + a_0$.

One correlation however will remain and will create some problems. This is
the *period–color* (PC) relation discovered by Eggen (1961, 1967). Its existence
is a mixed blessing. On one hand, we can achieve a partial substitution of
the uncertain color by the robust period; on the other hand, the regression
coefficients are difficult to determine, as the two parameters do not form an
good basis to split the luminosity dependence. As a result, determinations of
the coefficients $a$ are weakly constrained by "run-of-the-mill" systems that follow
the PC relation, but rather strongly depend on systems sometimes considered
as deviating from the norm. One must be careful with such systems: their $\dot{a}$

*priori* rejection could weaken our results whereas unwarranted retention could lead toward a wrong calibration.

We note that our calibration cannot be perfect because W UMa-type binaries are known to be very active, show large star-spots and strong chromospheric activity. Thus, we can expect random deviations of individual systems by a few 0.1 mags due to spots or some (perhaps systematic) deviations of colors, due to intrusion of chromospheric temperature increases into the outer photospheric layers of their atmospheres. We neglect these complications for the time being, but we should be aware of their existence.

Finally, we should answer a question which might have already been posed by the reader: Why not to use spectroscopic data? It should be simple to insert all parameters available from spectroscopic and photometric solutions in Eq. (1) or determine coefficients in this equation; why not to do that? The reason is primarily to avoid any model dependencies. For example, if we did not want to use the period to estimate the emitting area, $a^2 S(q, f)$, but instead used the spectroscopic value of the component separation, then $L \propto T_e^4 S(q, f) (K_1 + K_2)^2 / sin^2 i$. Since the velocity amplitudes, $K$, might depend on how one measures spectra, and the inclination, $i$, might depend on how one models the light curve, this is a risky route. Besides, only a handful of contact systems have really high-quality spectroscopic determinations and these are not those systems which delimit extremes of parameters, hence they are not the best for establishing parameter dependencies.

## 3. Calibration in practice

In selecting the initial sample of calibrating systems, only one principle was observed: attempt to avoid systematic errors. Thus only 18 systems were selected: 3 nearby W UMa systems with known parallaxes, 3 field systems with faint companions (to make sure that they can be placed on the Main Sequence) and 12 systems in high galactic latitude open clusters (to make sure that we do not include Milky Way interlopers). On the basis of this sample, so far 3 relations have been established (CAL1, BWC):

$$M_V = -2.38 \log P + 4.26 (B - V) + 0.28, \quad \sigma = 0.24 \qquad (2)$$

$$M_V = -4.43 \log P + 3.63 (V - I_C) - 0.31, \quad \sigma = 0.29 \qquad (3)$$

$$M_I = -3.23 \log P + 2.88 (V - I_C) - 0.03, \quad \sigma = 0.29 \qquad (4)$$

The period $P$ is in days and both colors are reddening-free.

Let us analyze one of these calibrations, perhaps the first one, giving $M_V$ in terms of $\log P$ and $B - V$. Basically, it defines a plane in in a cube, as in Figure 1. The ranges in parameters are relatively narrow at present so a plane suffices, but in future we might want to include some curvature terms. For the 18 system sample, the ranges are: $0.27 < P < 0.59$ day, $0.40 < B - V < 1.08$, and $0.46 < V - I_C < 1.19$ (the color $V - I_C$ was derived, not observed, see below) but, in the future, we might expect to extend the parameter ranges to perhaps as wide as $0.25 < P < 1.0$ day and $0.2 < B - V < 1.2$. Then, the respective contributions to $M_V$ would be $\Delta M_V \simeq 2.38 \times 0.6 \simeq 1.4$ and $\Delta M_V \simeq 4.26 \times 1.0 \simeq 4.3$. As we see, the color term dominates. However, the color dependence is shallower than on the Main Sequence and the period term cannot be entirely dropped as it is statistically significant (CAL1).

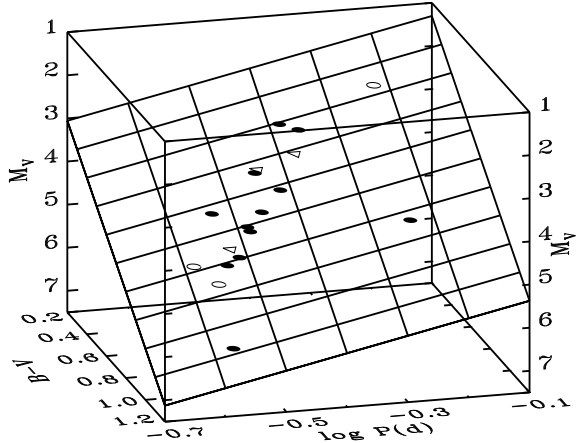

Figure 1. The $M_V(\log P, B-V)$ calibration shown as a cube. Note the period-color relation running along the diagonal and the strong dependence of the inclination of the calibration plane on one evolved system, V5 in NGC 188 (RHS of the cube, in the middle). Different symbols mark systems with $M_V$ estimated from trigonometric parallaxes (open ovals), from membership in high galactic altitude open clusters (filled ovals) or from visual MS companions (open triangles).

How well are the coefficients in Eqs. (2–4) determined? The answer is: Very poorly. Here traditional least-squares estimates fail entirely and the best way is to perform "bootstrap sampling" experiments (random selections with repetitions, see CAL1). They tend to be somewhat pessimistic for small samples, as ours, but give fair estimates of uncertainties. As is shown in Figure 2, the coefficients $a$ are strongly correlated and their uncertainties are large.

Although the regression coefficients are poorly determined, estimates of the absolute magnitudes are not as bad as one might initially think, and this is exactly because of the existence of correlations between the coefficients. In other words, if we make an error in one coefficients, the error in the other would be correlated in such a way as to compensate for the uncertainty. Our initial estimate was that the final uncertainty in predictions of absolute magnitudes would be at the level of about ±0.5 mag, but it seems that perhaps this is slightly too pessimistic, although useful as a ball park upper limit.

More extensive Monte Carlo experiments to estimate the absolute magnitude uncertainties were made for the $M_I(\log P, V-I_C)$ calibration which had been used for the Central Baade Window sample from OGLE (Udalski et al. 1994). Here the bootstrap samples of coefficients, determined as above for estimates of their uncertainties, were applied to randomly selected pairs $(P, V-I_C)$. This way, we can see directly how relatively large (but correlated) uncertainties in the coefficients $a$ would translate into deviations in $M_I$. The results are quite encouraging (Figure 3). The deviations are moderate and $\pm 1\sigma$ deviations tend to be as small as 0.2 mag. However, such small deviations are seen only within the strict range of periods used for the calibration. Outside this range, the deviations become much larger. Obviously, further efforts should go into extending the period baseline of the calibration.

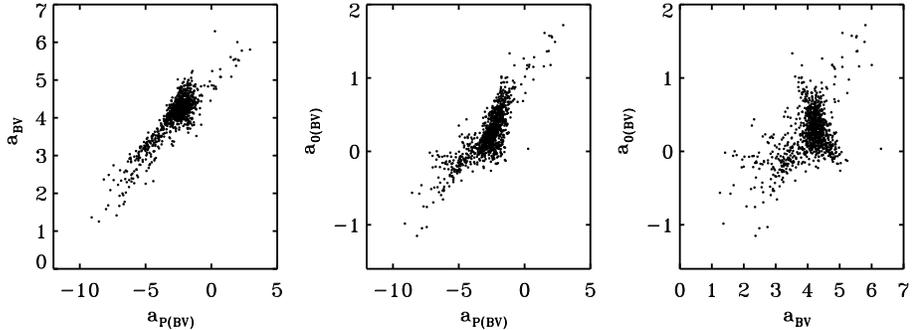

Figure 2. Results of 1000 bootstrap sampling experiments (CAL1) for the three coefficients of the relation: $M_V = a_{P(BV)} \log P + a_{BV}(B - V) + a_{0(BV)}$. Note large uncertainties of the coefficients and strong correlations between them.

## 4. Future work

### 4.1. The $V - I_C$ colors

It is not an accident that the section above was finished with the $M_I(\log P, V - I_C)$ calibration rather than that based on the $B - V$ color. The color index $V - I_C$ is becoming the most popular in searches for variable stars in open and globular clusters; it is also used in the gravitational micro-lensing experiments. Although the original motivation to use $I_C$ in place of $B$ might have been the higher sensitivity of the CCD detectors in the near infrared, the $V - I_C$ color index is more convenient than the $B - V$ because of its wider spectral base and much weaker sensitivity to metallicity variations (CAL2). But there is a problem: There are almost no $V - I_C$ data for bright field contact binaries.

Lack of the $V - I_C$ data for contact binaries is one of the most severe stumbling blocks in wide application of techniques described in this paper. Confronted with new discoveries of contact systems but not having proper standard data to compare them with, I decided to resort to a desperate move (CAL1, CAL2): I used $V - I_C$ colors for standard systems obtained through $B - V$ to $V - I_C$ transformations which had been established for Main Sequence stars (Bessell 1979, 1990). It must be stressed that this was really a stop-gap measure and that these preliminary results should be replaced by those based on real measurements of the $V - I_C$ colors. After all, contact binaries are known to be very active and there are indications from *ubvy* photometry (Rucinski & Kaluzny (1981) and Rucinski (1983)) that their colors might be affected by chromospheric phenomena. Besides, there exists an obvious effect that at any orbital phase, and especially at light maxima, we see a lot more of limb darkened areas on them than on normal, spherical stars, so the colors might be different.

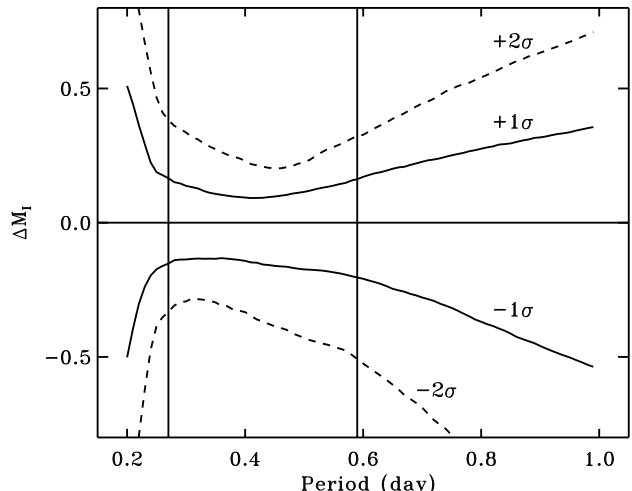

Figure 3. Ranges in $M_I$ resulting from uncertainties in calibration coefficients obtained by Monte Carlo experiments for the calibration $M_I(\log P, V - I_C)$. The vertical lines mark the range of periods for the 18 calibrating systems. This figure was not shown in the BWC paper for lack of space but certainly should be consulted to appreciate quality of distance determinations for contact systems visible toward the Galactic Bulge region (see Fig. 4).

Generally speaking, photometric information for contact binaries is in a very bad shape. This is in spite of large numbers of light curves obtained since first applications of photomultipliers in astronomy: There is probably no other type of variables being so frequently observed by anybody having access to a photomultiplier of a CCD. As the result, we have a lot of light curves but almost no color information. Maybe I am repetitious with my appeals made in different places and over so many years, but I will stress it again that *photometric data, in terms of colors and magnitudes are equally, or maybe even more important than light curves*. Among 562 W UMa-type systems in the most recent General Catalogue of Variable Stars, only some 50 or 60 have photometric data in standard systems. We badly need a new survey, similar to the old one in the $UBV$ by Eggen (1967), but giving several colors, at least the $V - I_C$ colors.

Having this shaky color information, only a very preliminary exploration of the metallicity dependence of the calibrations (CAL2) could be contemplated. This was possible thanks to discovery of contact binaries in very metal-poor globular clusters, NGC 5466 (Mateo et al. 1990) and NGC 4372 (Kaluzny & Krzeminski 1993). Both have $[Fe/H] < -2$. Although the systems are faint and difficult to observe, the metallicity baseline is wide enough to permits a rough estimate of the effect. Thus, for the $M_V$ calibration based on the observational, de-reddened color $B - V$, the correction term is: $\Delta M_V \simeq -0.3 \times [Fe/H]$. Comparison of the magnitude of this term with the model-atmosphere results on synthetic colors suggests that the largest effect in the calibration is through the color term; apparently, modifications involving stellar structure changes and reflecting in orbital periods are relatively small. This permits a prediction that the absolute magnitude calibration based on the $V - I_C$ color should be about 2.5 times less sensitive to metallicity variations than the one based on the $B - V$ color. This is again a good reason why the $V - I_C$ color should be more widely used.

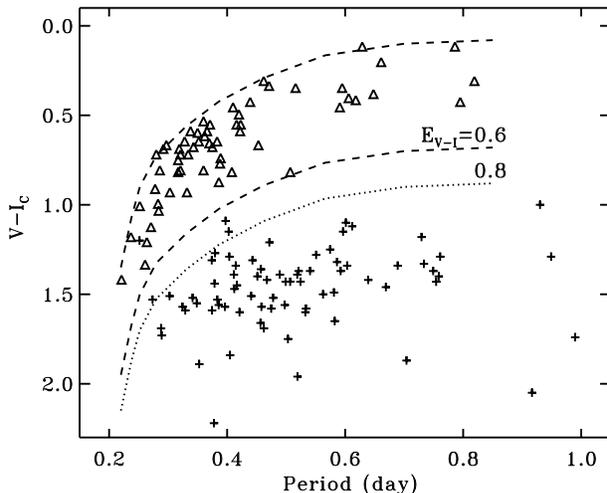

Figure 4. The period-color relation for contact systems discovered by the OGLE experiment in the Central Baade Window (crosses) compared with bright field systems (triangles), as tabulated by Mochnacki (1885), with $V - I_C$ colors transformed from the $B - V$ colors (BWC). The short-period, blue edge of the relation defines location of unevolved contact binaries showing no reddening (hand-drawn broken line).

### 4.2. Period range for standard systems

A look at the data for contact binaries discovered by OGLE in the Central Baade Window (BWC) shows that interpretation of the results is not straightforward (Figure 4). First, we would like to relate the new systems to the short-period, blue edge of the P–C relation of unevolved systems, but the latter is at present poorly defined and might contain a systematic problem due to the uncertain $B - V$ to $V - I_C$ transformation. Second, the data show combination of interstellar reddening and evolution for majority of the systems in the Baade Window. The reddening seems to be similar for all systems (i.e. mostly near foreground) and confined to about $0.6 < E_{V-I} < 0.8$[1], so that most of the right-downward shifts are due to evolution (but this should be of no consequence for our lumi-nosity estimates). And third, perhaps because of the admixture of many evolved systems, we see many relatively long-period contact binaries. This is because the long period systems are intrinsically brightest and are seen deep in space. But this is where our calibration is the weakest, as one can appreciate in Figure 3. For these systems random errors of the calibration are largest and we might easily have some large systematic problem there as well.

---

[1] The reddening information comes from several previous studies of RR Lyr stars in this field. There are about four times as many W UMa systems as RR Lyr stars in the OGLE sample.

There are relatively few long-period systems in the solar neighborhood and in clusters which we could use to extend the absolute magnitude calibration. It is not obvious whether they have been selected against exactly because of their long periods or they are genuinely rare. In "pencil beam" searches as OGLE such long period systems are expected to be over-represented exactly because of their brightness. In any case, such systems are really important if we want to use contact binaries to look deep into space.

### 4.3. Frequency of occurrence

The data coming from the micro-lensing experiments are probably the best material for statistical investigations that students of variable stars have ever had in their disposal. They are accompanied by excellent error analyses and have well known temporal selection-bias estimates. Equally precise, but less extensive in time, surveys of open clusters (Kaluzny & Rucinski (1993a), Rucinski & Kaluzny (1994)) have already shown that there are many more low-amplitude systems in the clusters than suggested by the field sample. This increase in numbers for small amplitudes is expected for random orbital inclinations. It suggests that the general field sample is heavily biased in favor of large-amplitude systems which were the first to be discovered in un-systematic searches of the sky.

Because of the addition of the low inclination systems, the open cluster sample indicates substantially more contact binaries than estimated before at about 1/1000 stars; perhaps by an order of magnitude more. But we must be careful: Some of these systems for low galactic altitude clusters (and this is majority) could be Milky Way interlopers. Here our calibration comes handy (CAL1): When systems deviating by more than $\pm 1$ magnitude are rejected as non-members, then we arrive at observed frequency of about 1/300, *which is more or less identical for the clusters and for their foreground/background field*. Correcting this for undetected low-inclination systems (by factor of 2), we obtain the space frequency of the W UMa systems around 1/150, which agrees very well with the X-ray based estimate of Mateo in Hut et al. (1992).

This result (CAL1) is based on a small sample of W UMa systems in just a few (but well searched) clusters and should be confirmed by further studies.

### 5. Conclusions

Contact binaries have many advantages as standard candles: There are many of them since they are Main Sequence stars and occur with high frequency. Also, apparently their luminosity calibrations are relatively insensitive to $[Fe/H]$. However, they are not as good standard candles as RR Lyr pulsating stars – but these appear only in old systems (older than about 12 Gyr); they are also not as bright as Cepheids – but these appear only in young systems and with very low spatial frequency.

Spot activity of late-type contact binaries is a problem which might be difficult to incorporate in the calibrations. Similarly, differences in mass-ratios between systems are at present entirely unaccounted for. In fact, indications that mass-ratios might correlate with the period might be one of selection effects which are known to plague any statistics based on bright field systems available for full, photometric and spectroscopic investigations. But mass-ratios are determinable really only using spectroscopic methods so that for faint systems the spread in $q$ will remain a problem.

In spite of all these precautions and qualifications, we think that we have now a good case to recommend these mysterious stars as reasonable standard candles. But why mysterious? In fact, we still do not know how they form, how are they built and how long do they last in the contact stage, a situation quite unusual for Main Sequence objects. But this is another story...

**Acknowledgments.** Support from the Natural Sciences and Engineering Council of Canada is acknowledged with gratitude.

**Discussion**

*Mario Mateo*: You note that there is a weak dependence of $M_V$ on $[Fe/H]$, yet there is a very strong effect in the period–color plot. How do you reconcile this?

*Slavek Rucinski*: This question relates to a fuller discussion presented in CAL2. For the metal-poor clusters NGC 5466 and NGC 4372, the color deviations are typically $\Delta(B-V) \simeq 0.2 - 0.3$, whereas the absolute magnitude deviations are typically $\Delta M_V \simeq 0.8 - 1.2$, which is consistent with the coefficient of the color term of about 4 (cf. Eq. (2)). To some extent this is a matter of semantics: 0.3 mag per unit of $[Fe/H]$ in $M_V$ is not much, but a deviation of 0.3 mag in $B-V$ sounds like quite much.

*Harvey Richer*: Were you able to estimate the distances to the globular clusters with your contact binaries? If so, did they agree with other estimates?

*Slavek Rucinski*: No, the process was done in reverse (CAL2). The distances were assumed in order to check if the contact binaries could possibly belong to these three globular clusters and some non-members were eliminated at this stage. Then the deviations were used to determine the metallicity-dependent term (Sec. 4.1.).

*David Turner*: If you have problems with estimating unique reddenings for W UMa systems, how do they make good standard candles?

*Slavek Rucinski*: If contact binaries really have colors of Main Sequence stars (as we have assumed) then any standard technique of reddening determination (such as two-color plots) should work. If they have different intrinsic colors, my hope is that sooner or later we will know them and will be able to establish a standard relation.

*Gene Milone*: Two comments: 1. We are obtaining $V$ and $I$ photometry of some of the systems in NGC 6791 and are investigating the effects of different metallicities on the light curve analysis solutions. 2. Linnell investigated the question of spots covering the cooler, more massive and larger component in the W-type contacts and was not able to replicate the light curves as well as the more traditional way of lower global $T_2$.

*Slavek Rucinski*: NGC 6791 where Janusz and I found a few contact binaries (Kaluzny & Rucinski 1993b) is emerging as one of the most interesting clusters. It is the oldest open cluster known, older by about 1Gyr than NGC 188; surprisingly, it is also very metal rich, perhaps with metallicity a few times solar, $[Fe/H] \simeq +0.5$ (Kaluzny & Rucinski 1995).

**References**

Bessell, M. S. 1979, PASP, 91, 589

Bessell, M. S. 1990, A&AS, 83, 357

Hut, P., McMillan, S., Goodman, J., Mateo, M., Phinney, E. S., Pryor, C., Richer, H. B., Verbunt, F., and Weinberg, M. 1992, PASP, 104, 981

Eggen, O. J. 1961, RoyObsB, No. 31

Eggen, O. J. 1967, MemRAS, 70, 111

Kaluzny, J. & Krzeminski, W. 1993, MNRAS, 264, 785

Kaluzny, J. & Rucinski, S. M. 1993a, in Blue Stragglers, ed. R. A. Saffer (San Francisco, ASP), ASP Conf.Ser. 53, 164

Kaluzny, J. & Rucinski, S. M. 1993b, MNRAS, 265, 34

Kaluzny, J. & Rucinski, S. M. 1995, A&A, in press

Mateo, M., Harris, H. C., Nemec, J. & OLszewski, E. W. 1990, AJ, 100, 469

Mochnacki, S. W. 1985, in *Interacting Binaries*, eds. P. P. Eggleton & J. E. Pringle (Reidel Publ. Co.), p.51

Rucinski, S. M. 1983, A&A, 127, 84

Rucinski, S. M. 1985, in Interacting Binary Stars, eds. J. Pringle, R. A. Wade, Cambridge Univ. Press, p. 85

Rucinski, S. M. 1992, AJ, 103, 960

Rucinski, S. M. 1993a, PASP, 105, 1433

Rucinski, S. M. 1993b, in The Realm of Interacting Binary Stars, eds. J. Sahade et al., Kluwer Acad. Pub., p. 111

Rucinski, S. M. 1994a, AJ, 107, 738

Rucinski, S. M. 1994b, PASP, 106, 462 (CAL1)

Rucinski, S. M. 1995a, PASP, 107, 648 (CAL2)

Rucinski, S. M. 1995b, ApJ, 446, L19

Rucinski, S. M. & Kaluzny, J. 1981, AcA, 31, 409

Rucinski, S. M. & Kaluzny, J. 1994, in Evolutionary Links in the Zoo of Interactive Binaries, edited by F.D'Antona et al., Mem. Soc. Astr. Ital., 65, 113

Udalski, A., Kubiak, M., Szymanski, M., Kaluzny, J., Mateo, M. & Krzeminski, W. 1994, AcA, 44, 317

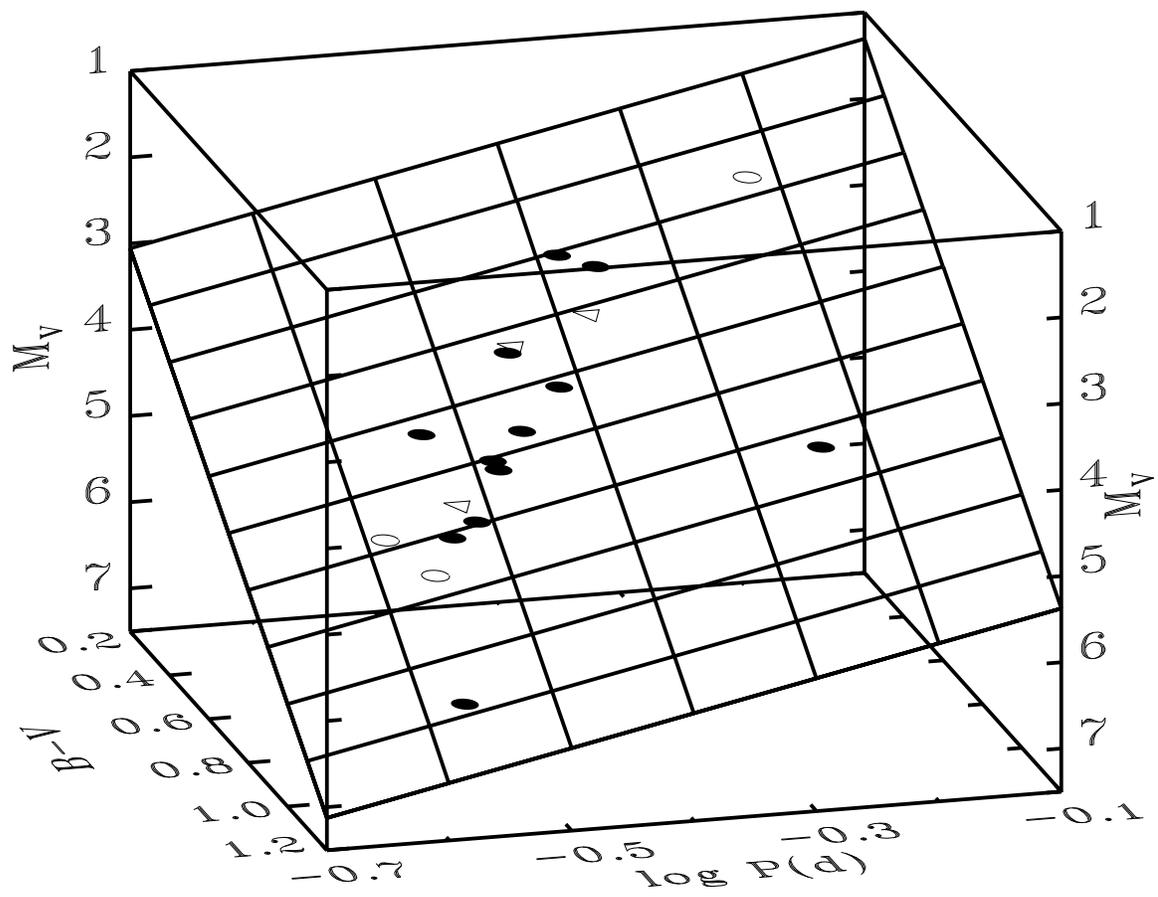

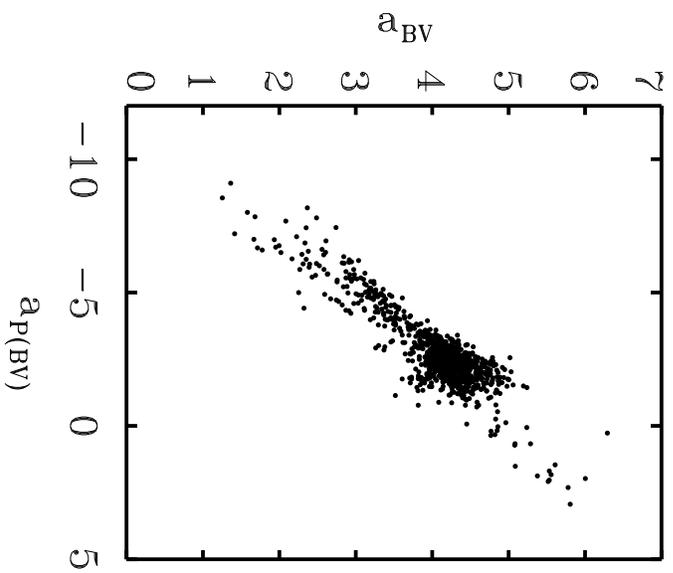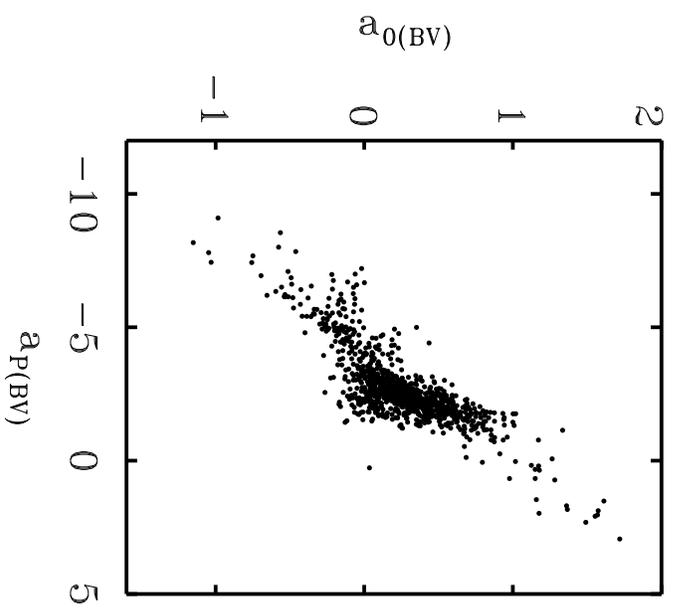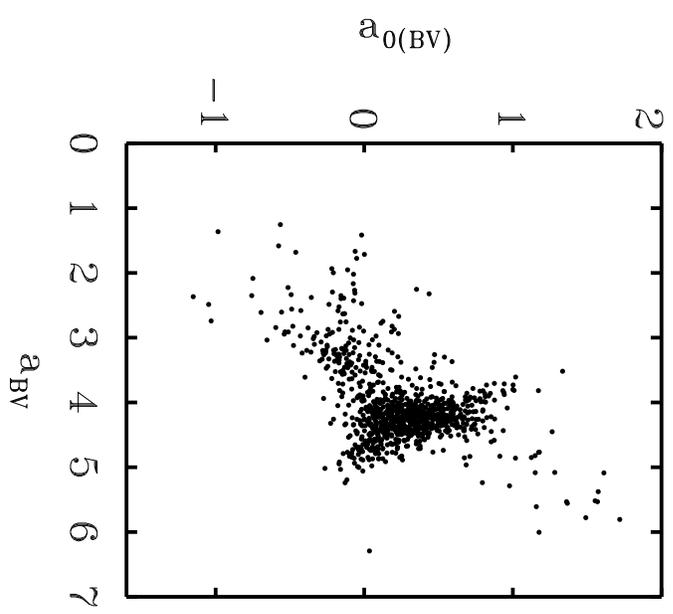

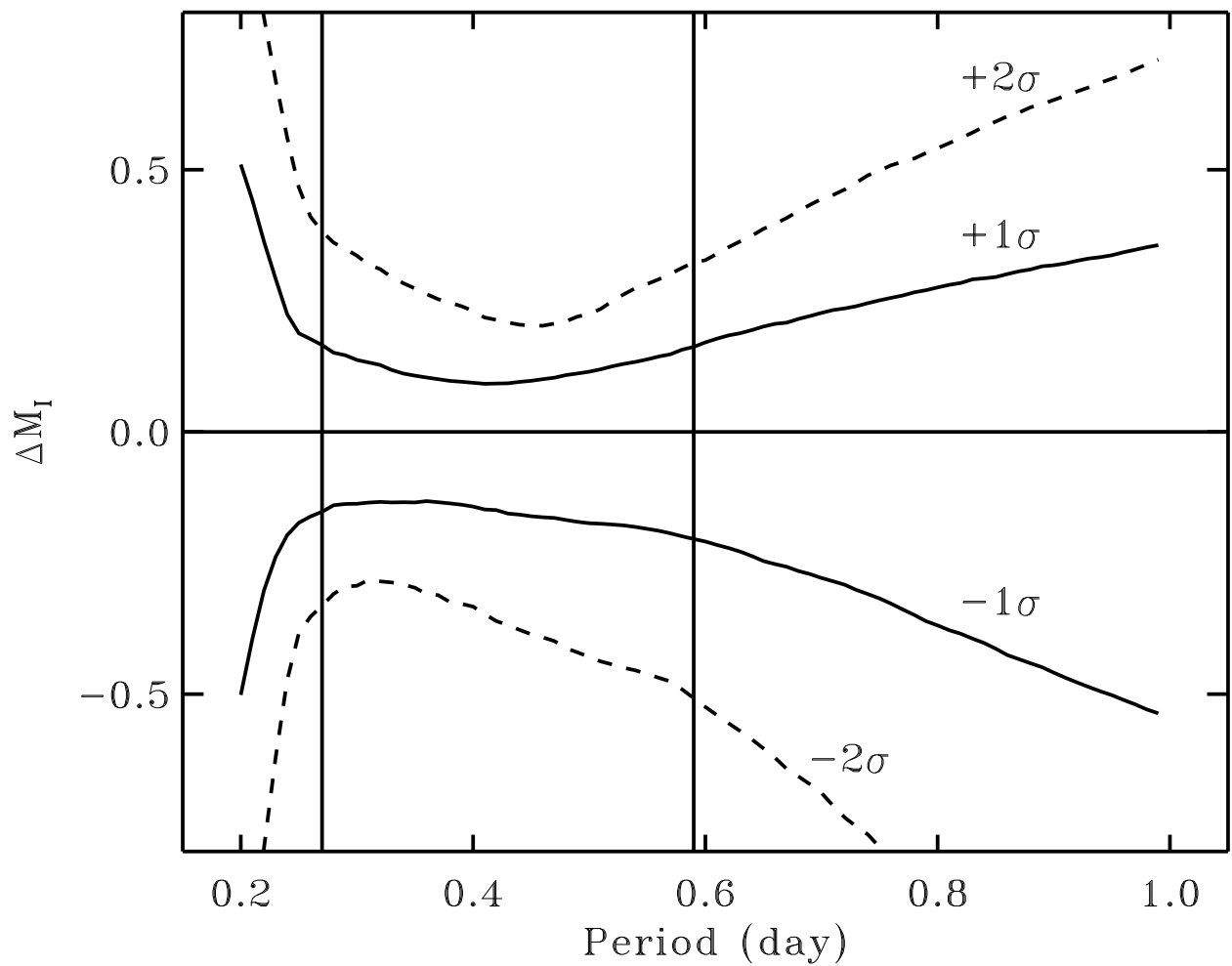

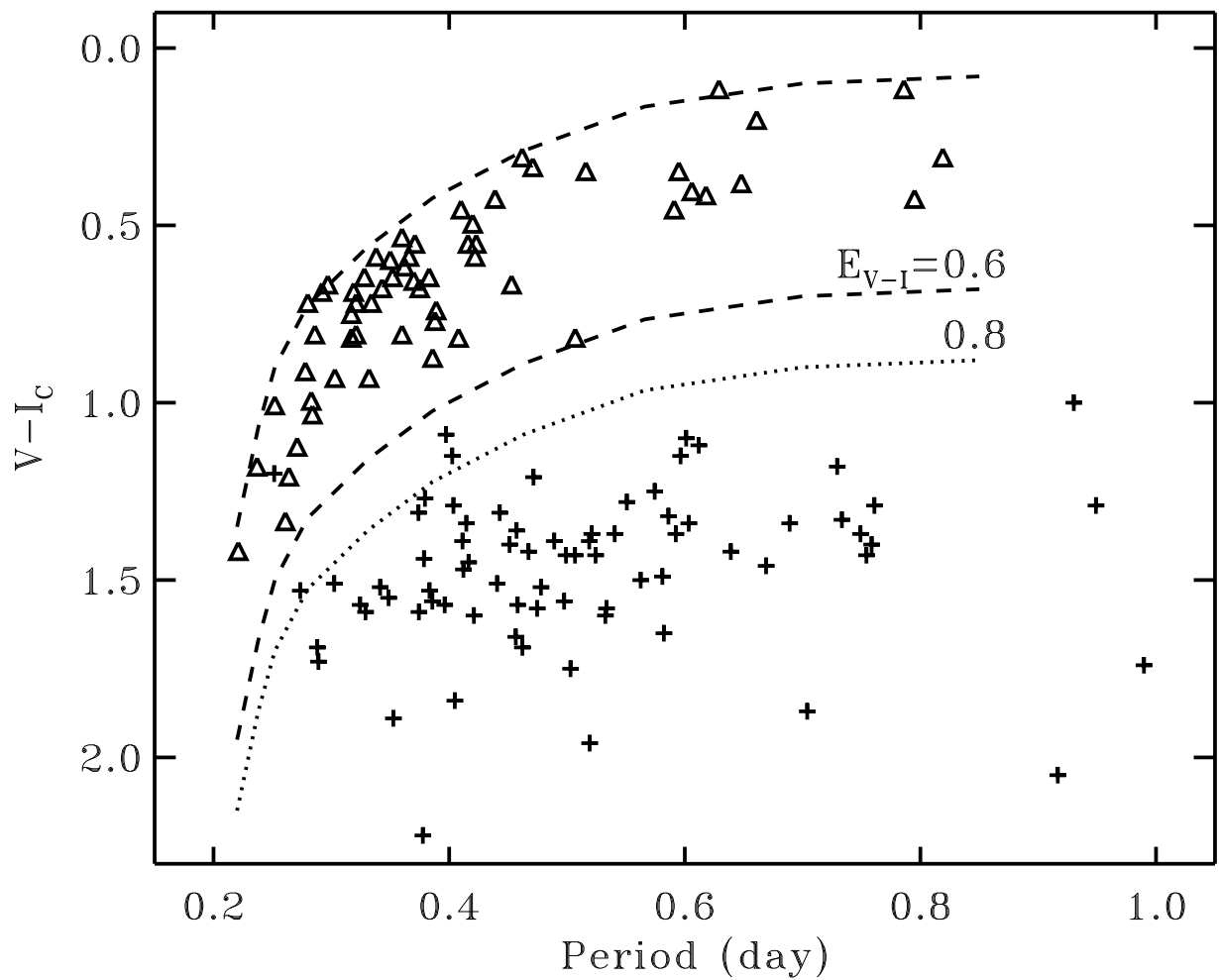